\documentstyle[12pt,epsfig]{article}
\begin{document}
\begin{center}
{\Large \bf Center vortices and confinement vs. screening}\\[.2in]

John M. Cornwall*\\
{\em Department of Physics and Astronomy\\
University of California, Los Angeles\\
Los Angeles, Ca 90095}\\[.2in]

{\bf Abstract}\\
\end{center}

\bigskip

\noindent Confinement in QCD is due to a condensate of thick vortices with fluxes in
the center of the gauge group (center vortices), as proposed long ago by the author and others.
It is well-known that such vortices lead to an area law for fundamental-representation Wilson loops, but what happens for screened ({\it e.g.}, adjoint) Wilson loops has been less clear, and problems have arisen
over the large-$N$ limit.  We study the adjoint and fundamental Wilson loops for
gauge group $SU(N)$ with general $N$, where there are $N-1$ distinct
vortices, whose properties (including collective coordinates and actions) we discuss.  In $d=2$ we construct a center-vortex model by hand so that it has a smooth large-$N$ limit of fundamental-representation Wilson loops and find, as expected, confinement.  Extending an earlier work by the author, we construct the adjoint Wilson-loop potential in a related model for all $N$, as an expansion in powers of $\rho/M^2$, where $\rho$ is the vortex density per unit area and $M$ is the gauge-boson mass (inverse vortex size) and find, as expected, screening.  (This is, in fact, unexpected in $d=2$ QCD.)
The leading term of the adjoint potential shows a roughly linear regime followed by string breaking when the potential energy is about $2M$.  This leading potential is a universal ($N$-independent at fixed $K_F$) function of the type $(K_F/M)U(MR)$, where $R$ is the spacelike dimension of a rectangular adjoint Wilson loop and $K_F$ is the \\[.2in]
\footnoterule
\noindent *E-mail address:  cornwall@physics.ucla.edu\\[.1in]
\noindent UCLA/97/TEP/30  \mbox{} \hfill December, 1997
\newpage
\noindent fundamental string tension.  The linear-regime slope is not necessarily related to $K_F$ by Casimir eigenvalue ratios.  We show that in $d=2$ the dilute vortex model is essentially equivalent to true $d=2$ QCD in the fundamental representation, but that this is not so for the adjoint representations; arguments to the contrary are based on illegal cumulant expansions which fail to represent the necessary periodicity of the Wilson loop in the vortex flux. Most or all of these arguments  are expected to hold for $d=3,4$ as well, but we cannot calculate explicitly in these dimensions (a proposal is made for another sort of approximation in $d=3$, using earlier work where d=3 vortices are mapped onto a scalar field theory.).\\[.4in]
\begin{center}
{\bf I.  INTRODUCTION}
\end{center}

There are many viewpoints concerning the mechanism of confinement in QCD.  The problem here is to make a confinement proposal which is sufficiently specific ({\it e.g.}, not dependent on choice of gauge) to allow for good tests of its correctness.  Recently there has been very considerable progress, both in lattice gauge theory \cite{t93,tk} and lattice gauge simulations \cite{tk,fgo}, in verifying the center-vortex picture of confinement \cite{c79,th79,mp,no,y80}, at least for the group $SU(2)$ in the fundamental representation.  The essence of the center-vortex picture is the existence of a condensate of closed magnetic sheets (in $d=4$) or closed magnetic strings (in $d=3$) which have a finite\footnote{Therefore on the lattice these vortices are infinitely spread out, in the limit of zero lattice spacing.  There are also vortices of a single lattice spacing in thickness, which have infinite action in the continuum limit and are completely suppressed.  The lattice problem is to characterize these spread-out vortices.  We will only consider the continuum picture of vortices \cite{c79}.} thickness $\sim M^{-1}$, where $M$ is the gauge-boson induced mass \cite{c82}.  These vortices carry (color) magnetic fields and have magnetic fluxes which lie in the center of the gauge group.  They form a condensate because their entropy (per unit size) is larger than their action.  Their continuum description is essentially that of the Abelian Nielsen-Olesen vortex, with some modifications to account for the difference in mass generation mechanisms between QCD (no gauge symmetry breaking) and the Abelian Higgs model\footnote{Callan, Dashen, and Gross \cite{cdg} studied confinement in the $d=2$ Abelian Higgs model for matter fields with fractional charge; except for details of group structure, this model captures the essence of the center-vortex picture of confinement.  However, it was only later that the precise connection of thick vortices and confinement was made in non-Abelian gauge theories.}.  The gauge potential has both a short-range part and a long-range pure-gauge part corresponding to a gauge transformation which is singular along the center of the vortex; the field strength is non-singular and purely short-range.  An area law for a large (compared to $M^{-1}$) fundamental Wilson loop arises because the long-range pure-gauge part of the vortices with non-zero Gaussian linking number contributes a factor like $\exp (2\pi iK/N)$ to the Wilson loop, for gauge group $SU(N)$.  Here $K$ is an integer which is the product of the linking number and the magnetic flux of the vortex. The vortices are linked randomly, and an average over all vortex linking numbers yields an area law \cite{c79}.

Striking confirmation for this picture of confinement has been found in recent lattice studies of the fundamental representation of $SU(2)$.  The simple procedure is to replace, for a given configuration of lattice gauge potentials, the true Wilson loop by its sign (corresponding to the vortex factor
$\exp (i\pi J)$ for link number $J$ in $SU(2)$).  To the numerical accuracy of the lattice computations, the string tension so found is precisely the same as for the full theory \cite{tk,fgo}.  The only difference is found at distances comparable to the physical scale of the theory (that is, $M^{-1}$), where the finite thickness of the vortices and other perimeter-law effects begin to show up.  

However, present-day theory and numerics of the center-vortex mechanism leave a number of questions unanswered, and we will address them in this paper.  In the first place, no work is known to the author which studies, either theoretically or computationally, the nature of the vortex condensate and the behavior of the fundamental Wilson loop for generic $N$ with gauge group $SU(N)$.  (The case of $SU(3)$ was studied in Ref. \cite{c79}).  Nor has there been much work on the particularly important problem of the behavior of other representations of the Wilson loop, in particular the adjoint Wilson loop.  There is an old work of the author \cite{c83}, obscure and little-known, which discusses this question; we refine and extend that work here.  In addition there is some recent speculation by Faber {\it et al.} on the adjoint potential \cite{fgo2} and the possibility of so-called Casimir scaling, which holds that the ratio of string tensions in the fundamental and adjoint representations is simply the ratio of quadratic Casimir eigenvalues for these two representations.  Here one must understand, of course, that the adjoint string tension cannot persist to indefinitely large distances, since it can always be screened by gluon-pair formation.  Ultimately the string must break (see, {\it e.g.}, Ref. \cite{cb} for an early lattice calculation of this effect; references to other lattice calculations of a similar type can be found in Ref. \cite{fgo2}).  

Casimir scaling is certainly true for $d=2$ QCD, but this theory is in some respects a misleading model, for it shows an area law to all distances in the adjoint representation (simply because there are no gluonic degrees of freedom to screen the Wilson loop).  Greensite and Halpern \cite{gh} have given a large-$N$ argument to the effect that the usual large-$N$ factorization property (from which follows Casimir scaling) requires an adjoint area law in $d=3,4$ as well. This is fatal to the center-vortex picture, where in these dimensions the adjoint representation is certainly screened and not confining.  At large $N$, Casimir scaling simply says that the adjoint string tension is twice the fundamental string tension, which can be interpreted as the presence of two fundamental strings between an adjoint ``quark" and its antiparticle.

An essential part of the present work is to show that there is no large-$N$ contradiction in the center-vortex model between fundamental-representation area laws and adjoint perimeter laws (although this is superficially in contradiction to conventional large-$N$ factorization; the resolution of the contradiction is discussed in Section V.).  The fundamental and adjoint representations must be treated differently, unlike the case of $d=2$ QCD, and only the fundamental representation shows an area law.  We do not find Casimir scaling for the adjoint (breakable) string tension, nor do we find conventional large-$N$ factorization properties. This agrees with Greensite and Halpern's argument that the center-vortex model is inconsistent with such factorization.   Although we will spend considerable time on a $d=2$ vortex model\footnote{See also the work of Smilga \cite{smil}, which invokes vortices in $d=2$ gauge theories with fermions.}, it must be understood that this model is by no means the same as $d=2$ QCD, which does have confinement and Casimir scaling in all representations.  It is indeed true, as we will show, that the $d=2$ vortex model and QCD can be made to look the same in the fundamental representation, but not in the adjoint, for which we will calculate the first term in a density expansion of the adjoint potential and find \cite{c83} a potential which is approximately linear for intermediate distances, but breaks when enough energy is stored in the potential.  We emphasize the $d=2$ vortex picture not because we believe that 2 is the relevant dimension for the model; in fact it is not.
The relevant dimensions are 3,4, but it is more difficult to make explicit calculations with the vortex picture in these dimensions.\footnote{In another publication we will estimate the $d=3$ adjoint potential from vortices, using \cite{cy} a description of the vortex condensate in terms of a scalar field, which is appropriate for this dimension.}  

It is instructive to revisit some simple but flawed arguments (see, {\it e.g.}, \cite{feyn,c82}; these authors explicitly point out the flaw) showing that confinement comes essentially from a gluonic mass gap.  These arguments also lead (incorrectly, since they do not apply at $d=2$) to Casimir scaling and an area law for the adjoint representation.  The flaw is analogous to a truncated expansion of a periodic function like $\cos(\phi)$ in $\phi$, saving only second-order terms.  But with such a truncation periodicity is lost, and all sorts of bizarre effects can arise from the non-periodicity of $\phi$ itself.  Precisely this sort of thing comes up in the QCD arguments, as we will discuss later on.  The point is that the potential coming from the Wilson loop is periodic in the long-range pure-gauge vortex flux, with period $2\pi $ for the fundamental representation and period $2\pi /N$ for the adjoint.
In the first case the potential depends on the flux $\Phi$ as something like $\cos(\Phi)-1$, and in the second case it is like $\cos(N\Phi)-1$.  Since the flux is an integral multiple of $2\pi/N$ the long-range pure-gauge part of the vortex contributes nothing to the adjoint potential as expected.  But various simple approximations, like expanding to second order in the flux, completely obscure this fact.  
\bigskip

\begin{center}
{\bf II. THE CENTER VORTEX PICTURE}
\end{center}

First we review the general center-vortex picture, then we give some details of the flux matrices describing $SU(N)$ vortices.

\bigskip

\begin{center}
{\bf A.  Center Vortices and the Gluon Mass}
\end{center}

Because of infrared instability in $d=3,4$, QCD in these dimensions generates a dynamical mass for the gluon, proportional to the invariant mass scale of the theory \cite{c82,chk,c97}.  The existence of this mass is discovered by studying the Schwinger-Dyson equations of QCD, in a framework \cite{c82,chk} which insures gauge invariance, and noting that ``wrong-sign" effects associated with infrared instability lead only to solutions with a massive gluon.  It is necessary to include longitudinally-coupled massless scalars in the Green's functions, which play a role rather like Goldstone excitations.  Like Goldstone particles, these massless scalars do not appear explicitly in the S-matrix.  However, they play a crucial role in confinement.  Unlike Goldstone particles, they do not signal any sort of breakdown of local gauge symmetry, which is completely preserved.

An effective theory, accurate in the infrared, for describing the gluon mass is \cite{c82} the gauged non-linear sigma model.  This theory is useful in any dimension $d$; the corresponding (Euclidean) action $I$ is:
\begin{equation}
I=\int d^dx \{ \frac{-1}{2}Tr [G^a_{ij}(x)]^2 -M^2Tr [D_iU]^2 \}.  
\end{equation}
The gauge potentials are described by the usual anti-Hermitean matrices
$(\lambda^a/2i)A_i(x)$, and $D_i$ is the covariant derivative.  The $N\times N$
unitary matrix $U$ describes the non-linear sigma model fields.  Note that the gauged non-linear sigma model is locally gauge-invariant.  To use the effective action (1), one solves the equations of motion for $U$ in terms of the gauge potentials and substitutes the result in the equations for the gauge potential.  One then finds the above-mentioned massless scalar modes.  There also may be solutions for $U$ containing terms not dependent on the gauge potentials; this possibility is important for vortices.  The effective action (1) is not renormalizable, and breaks down in the ultraviolet.  This breakdown simply reflects the fact that the gluon mass $M$ is taken to be constant in (1), while the solutions of the Schwinger-Dyson equations insist that the mass be a function of momentum $p$, vanishing at large $p^2$ \cite{c82}.  In fact, Lavelle \cite{la} has shown that in $d=3,4$ the mass vanishes (modulo logarithms) like $1/p^2$ times the condensate expectation value $\langle G^2 \rangle $.

For constant $M$ the effective action (1) has numerous solitonic solutions.
Vortex solutions exist in all dimensions, with the vortex co-dimension fixed at 2.  This means that the vortex is describable in terms of an arbitrary closed
$d-2$-dimensional surface, which reduces to a point in $d=2$.  Most of what we can explicitly do with the center-vortex model will be in $d=2$, so we write the solution in that dimension, for a vortex centered at the origin:
\begin{equation}                                           
A_j(x;J)=(2\pi Q_J/ig)\epsilon_{jk}\partial_k\{\Delta_M(x)-\Delta_0(x)\}.           \end{equation}                 
Here $\Delta_{M,0}$ are free propagators for mass $M,0$, and $Q_J$ is a generator of $SU(N)$ such that $\exp(2\pi iQ_J)$ is in the center $Z_N$.  We choose the integer $J$ such that the corresponding element of the center is
$\exp(2\pi iJ /N)$.
The reason \cite{c79} for this requirement is that if a gluon is transported around a  large circle containing the vortex at the origin, it suffers a gauge transformation $\exp(2\pi iQ_J)$,
which must leave the gluon field unchanged.  This and other long-range effects come from the massless term in (2), which is a pure singular gauge term, expressible as the gradient of the polar angle.  Its singularity at the origin is exactly cancelled by the $\Delta_M$ term, and the field strength is short-ranged.  The solution (2) is immediately generalized by applying space-time translations and global group rotations, yielding the vortex as described in terms of its zero-mode collective coordinates:
\begin{equation}                             
A_j(x;a,u;J)= V^{\dag}(u)A_j(x-a;J)V(u).                            
\end{equation} 
Here $V(u)$ is the (fundamental) representative of group element $u$.  Of course, only a coset of the full group corresponds to true zero modes.  We will describe this coset in connection with a description of the flux matrices
$Q_J$,  and in the Appendix.

One readily calculates the contribution of the long-range pure-gauge part of a single vortex of flux $J$ to a simple (non-self-intersecting) Wilson loop $W$ in the {\em fundamental} representation.  Let the vortex be centered at position $a$ (see equation (3)); then
\begin{equation}                                                  
TrP\exp g\oint dx_iA_i(x)=Tr\exp (2\pi iQ_J)\Theta_W(a)+Tr1(1-\Theta_W(a)).
\end{equation}
Here $\Theta_W(a)$ is the characteristic function of the Wilson loop, that is, it is unity if the vortex lies inside the Wilson loop and zero if it lies outside.  If the loop is not simple, and wraps $K$ times around $a$, then $Q_J$ is replaced by $KQ_J$.  In view of $\exp (2\pi i Q_J)=\exp (2\pi iJ/N)$, the Wilson loop for a single long-range vortex is just the exponential of $2\pi iJ/N$ times the Gauss linking number of the loop and the center of the vortex, a situation which is true in any number of dimensions \cite{c79}.  An area law for the fundamental Wilson loop follows immediately, as discussed in many places (see \cite{cdg,c79,t93,tk}).In the {\em adjoint} representation, $J/N$ in the above is replaced by $J$, so the long-range pure-gauge part of the vortices is invisible to the adjoint Wilson loop, as is well-known.

At this point it is not clear how to proceed to the large-$N$ limit.  The reason is (see Sec. IIB immediately below and the Appendix) the action $I(N,J)$ of a vortex of flux $J$ depends on $J,N$ in such a way that when $J\sim N$ the action is very large compared to the action for fixed $J$ as $N\rightarrow \infty$.  However, in computing the statistical weights for vortices it is not just
$\exp (-I)$ which is important; one must also calculate the factors coming from group collective coordinates and other entropic effects.  In the Appendix we show that in a $d=2$ center-vortex model, which is constructed by hand, it is indeed possible to find a smooth large-$N$ limit when group collective coordinates are accounted for.  These tend to counter the exponential of the action.  To simplify further explanation of the large-$N$ center-vortex picture, we will simply assume that all vortices of whatever flux have the same free energy.  This may or may not be literally true, but there is no qualitative difference in the discussion of large $N$ whether it is or not.  We will find confinement for the fundamental representation and screening for the adjoint.

\bigskip 

\begin{center}
{\bf B.  Vortex Flux Matrices}
\end{center}

For future use we need some properties of the vortex flux matrices $Q_J$.
First introduce the traceless matrices $Q_i,\;i=1\dots N$:
\begin{equation}                                     
Q_i= diag(1/N,\dots 1/N, -1+1/N, 1/N,\dots 1/N)
\end{equation}
where $-1+1/N$ is in the $i$th position.  Of course, these are not independent; the sum over all $i$ is zero, and any $i$ can be reached from any other by a group rotation.  Any of these matrices is associated with a flux of $1/N$, in the sense
\begin{equation}                                       
\exp(2\pi iQ_j)=\exp (2\pi i/N).
\end{equation}
One easily checks that the matrices $Q_i+Q_j+Q_k+\dots $, with $i\neq j \neq k
\neq \dots$, has flux $J/N$ if there are $J$ terms in the sum; that is, 
\begin{equation}                                      
\exp (2\pi i(Q_i+Q_j+Q_k+\dots ))=\exp (2\pi iJ/N). 
\end{equation}
For $1\leq J \leq [N/2]$, where $[\cdot ]$ indicates the integral part, we can choose in any convenient way one representative of the above matrices as $Q_J$, the representative of flux $J$ in the sense of equation (7).  For larger values of $J$ one uses instead the matrices $Q_{N-J}\equiv Q_{-J}$, which are anti-vortices with charge $-J$.

The addition of several $Q_i$ to get a vortex of different charge (including the vanishing of the addition of $N$ of them) has a physical interpretation \cite{c79,c94}.  In $d=3$, vortex strings (surfaces in $d=4$) can merge or split at a point (line) with conservation of vortex charge.  In this way a string  (surface) network is formed.  If $N$ unit vortices meet, they can annihilate.  Every intersection point (line) is associated with a QCD sphaleron (sphaleronic world line), carrying a change of topological charges which is quantized in units of $1/N$, like the flux itself.  This has been explicitly illustrated in $SU(2)$ \cite{ct86}.

We can now see the structure of the coset needed to describe the group zero modes.  Any $Q_J$ has $J$ diagonal elements which are all equal, and $N-J$ diagonal elements also all equal to each other, but not equal to the first group.  Therefore it is invariant under $SU(N-J)\otimes SU(J)\otimes U(1)$, and the collective-coordinate coset is $SU(N)$ divided by the above.  More details are given in the Appendix.

It is clear from Section IIA that the vortex action depends on $TrQ_J^2$, so we record it:
\begin{equation}                               
TrQ_J^2=\frac{J(N-J)}{N}.
\end{equation}
Then every vortex has a different action\footnote{If one tries to calculate the action from (1) and (2), one finds a short-distance logarithmic divergence associated with the constant mass.  This is cured if one recognizes that the mass actually vanishes at short distances, as discussed above.}, which at first glance makes it difficult to see how the various elements of the center group are treated equally.  A related problem is treated in the Appendix, where it is shown how the $d=2$ vortex model can be tuned to have approximate equality of free energies for vortices with $J\sim N$; it is hoped that this equality of free energies emerges dynamically in $d=3,4$. 

Finally, we note that in the adjoint representation every $Q_J$ is a diagonal matrix, with $(N-1)^2$ eigenvalues of $0$, $N-1$ eigenvalues of +1, and $N-1$ eigenvalues of -1.

We are now ready to calculate Wilson loops, both fundamental and adjoint, with the usual machinery of dilute-vortex expansions.  However, before carrying this out it is perhaps instructive to discuss some flawed arguments which seem to contradict what we have said about center vortices and their role in confinement.

\bigskip

\begin{center}
{\bf III.  GOOD AND BAD CUMULANT EXPANSIONS}
\end{center}

We exhume some old and rather rough arguments concerning the mechanism of confinement \cite{feyn,c82} in $d=3,4$.  In any representation $R$ of $SU(N)$, we make two approximations to the Wilson-loop expectation value.  First, we convert the line integral to a surface integral, but we use the usual Abelian form of Stokes' theorem rather than the correct non-Abelian form; second, we make a cumulant expansion, saving only the first non-vanishing term, which is equivalent to assuming a Gaussian distribution of gauge potentials.  The result is
\begin{eqnarray}                                                
\langle W_R \rangle \equiv \langle TrP\exp [ig\oint dx^iT^aA_i^a]\rangle \\
\simeq D_R\exp [-\frac{g^2C_R}{2(N^2-1)} \int \int d\sigma_{ij}d\sigma'_{kl}
\langle G(x)^a_{ij}G(x')^a_{kl} \rangle]\nonumber.
\end{eqnarray}  
Here $D_R$ is the dimension of representation $R$, and $C_R$ is the quadratic Casimir eigenvalue for this representation, given by
\begin{equation}
C_R\delta_{ab}=\frac{Tr(T^aT^b)(N^2-1)}{D_R}.                 
\end{equation}
If the field strengths are short-ranged, so that $\langle G(x)G(x')\rangle
\sim \exp(-M|x-x'|)$ at large distances, one easily sees that there is an area law for Wilson loops large compared to $M^{-1}$.
Evidently this area law shows Casimir scaling, and it also scales correctly at large $N$, since $\langle (G^a)^2 \rangle \sim N^2-1$ and $C_R \sim N,\;g^2 \sim 1/N$.
In fact, (9) is quite correct for $d=2$ QCD, where Wilson loops of any representation are calculated using free gluon propagators and there are no gluonic self-interactions.  But it certainly cannot be correct in $d=3,4$ because it gives an area law for the adjoint (and other $N$-ality 0) representations.  The argument given here completely ignores the vital role of long-range pure-gauge parts of the potential, which is in fact the secret of confinement, as discussed above.  If one did not know about these long-range parts, one could not understand confinement, since the original (line-integral) form of the Wilson loop would surely give only a perimeter law, if all the gauge potentials were short-ranged.  About the only thing that the above argument really shows is that long-ranged gauge field strengths (as in QED in $d=4$) cannot confine.

From the point of view of understanding the adjoint potential coming from center vortices, this argument is an example of a bad cumulant expansion, which ignores a fundamental requirement of periodicity in the vortex flux.  Later we will see how this bad cumulant expansion can be found from an illegal truncation of the dilute center-vortex calculations.

There is a closely-related illegal argument which recognizes the existence of vortices with quantized flux.  Consider the expectation value of a large fundamental Wilson loop in $SU(2)$; according to the above we have
\begin{equation}
\langle W \rangle = \langle e^{i\pi J} \rangle.
\end{equation} 
Here $J=\sum J_i$ is the sum of the linking numbers of all the vortices which link with the Wilson loop.  Since $J$ is the sum of a large number of independent random numbers, one might be tempted to use the central-limit theorem and write
\begin{equation}
\langle W \rangle = \exp(-\frac{1}{2}\pi^2\langle J^2 \rangle)
\end{equation}
and then argue that $\langle J^2 \rangle $ is proportional to the number $N$ of vortices linked to the loop.  Given an areal density $\rho$ of vortices, we have
$N=\rho A$, where $A$ is the area of the loop, and an area law follows.  But consider the same argument for an adjoint Wilson loop; it surely is wrong to say that
\begin{equation}
\langle e^{2\pi iJ} \rangle = \exp (-\frac{1}{2}(2\pi )^2\langle J^2 \rangle).
\end{equation}
since $\exp (2\pi iJ)=1$ always.

We are now ready to use the center-vortex picture correctly.

\begin{center}
{\bf IV.  DILUTE-VORTEX EXPANSION OF THE ADJOINT VORTEX POTENTIAL}
\end{center}

The dilute-vortex expansion is of conventional type.  Given a set of solitonic fields
$\{\phi_J\}$ and their collective coordinates $\{ c_J\}$, the expectation value of any operator $O\{\phi (x)\}$ is approximated by the leading semi-classical term:
\begin{equation}
\langle O\{ \phi (x)\} \rangle = Z^{-1}{\large [ \sum}_K \frac{1}{K!}\sum_1
\dots \sum_K O\{ \sum_{J=1}^K \phi(x;c_J)\} {\large ]}.
\end{equation}
The partition function $Z$ is the same sum with $O\equiv 1$.  The sum over $K$ is a sum over sectors with $K$ solitons each.  The sums $\sum_1\dots \sum_K$ are each a sum over all the collective coordinates $c_J$, which include translations and group coordinates in $d=2$; in $d=3,4$ (and this is what makes the calculations there hard) there are sums over internal degrees of freedom of the strings or surfaces.  Implicit in the sum over multiple vortices are combinatoric factors appropriate to the different vortex charges.

In any number of dimensions, the collective-coordinate normalization factors lead to a specific dimensionful number, the density of vortices per unit area.  We call this density $\rho$; it has dimensions of mass squared.  This is evidently true in $d=2$.  In $d=3$, this density is simply the number of vortex string crossings (each crossing with unit weight) of any large rectangular area divided by the area, with an analogous definition in $d=4$.  Ultimately this density is converted to dimensionless form by dividing by the only available scale\footnote{In the dilute-vortex approximation, the coupling constant $g$ does not appear.} M, so the $K$-vortex sector is associated with a factor
$\rho /M^2\equiv \epsilon$.  This is essentially the vortex density multiplied by the vortex cross-sectional area.  We expect $\epsilon$ to be fairly small, since if vortices get too close together there is both an action penalty and an entropy penalty, but we do not know what the value of $\epsilon$ is.  Right now, it is just a hope that it is small enough to be a decent expansion parameter.  When we come to the adjoint potential we will calculate a part of the first-order term in $\epsilon$.

Everything we can do explicitly will be in two dimensions, where we can write the collective-coordinate integral for a single vortex of charge $J$ as
\begin{equation}
\sum_1 = \frac{\rho}{N-1} \int d^2a \int d(u)
\end{equation}
where the group integral is normalized to unity (so that it does not matter whether we integrate over the whole group or only over the coset appropriate to a given vortex).  Note that we assume no dependence in the collective-coordinate integral on the index $J$, as discussed in Sec.II.  The dependence $1/(N-1)$ in (15) merely reflects the fact that there is a total of $N-1$ different vortex types.         

As we have discussed earlier, it is essential to respect the periodicity of Wilson loop expectation values in the vortex flux.  This we can do straightforwardly for the $\epsilon^0$ term of the cluster expansion, but it seems to be much harder to do for higher-order terms, which we can discuss only qualitatively.  If the $\epsilon^0$ term itself is expanded in powers of the vortex gauge potential, with only the leading terms saved, one obtains results equivalent to confinement for the adjoint representation.

\bigskip

\begin{center}
{\bf A.  Fundamental Representation}
\end{center}

We will only consider the case of large Wilson loops, where to find the area law one need keep only the long-range pure-gauge part of the vortex (equation (2)).
In this case the calculation differs only in group-theoretic details from the Abelian Higgs-model version \cite{cdg}.  It is evident that 
\begin{equation}
Z= \exp (\rho V)
\end{equation}
where $V$ is the volume of the two-dimensional space.  Consider first $SU(2)$, where there is only one type of vortex.  The Wilson loop factors into products of the type in equation (4), with $\exp i\pi$ if the collective coordinate of the vortex is inside the Wilson loop, and unity otherwise. An elementary exercise in dividing by $Z$ yields just the Callan-Dashen-Gross result for the fundamental string tension $K_F$:
\begin{equation}
\langle W \rangle = e^{-2\rho A};\;K_F = 2\rho.
\end{equation}
Here $A$ is the area of the Wilson loop.

One might note here, by the way, that this result implies that the vortices obey Poisson, not Gaussian, statistics, that is, $\langle W\rangle$ can be written as
\begin{equation} 
\langle W \rangle = \langle \exp(i\pi L) \rangle
\end{equation} 
where on the right-hand side the expectation value is defined in terms of the Poisson probability
\begin{equation}
P(L)=\frac{\bar{L}^Le^{-\bar{L}}}{L!};\;\bar{L}=\rho A.
\end{equation}
Evidently this gives no area law for the adjoint representation.

For $SU(3)$, which has a vortex and an anti-vortex, the answer \cite{c79} is
\begin{equation}
\langle W \rangle = \exp [-\rho A(1-\cos(2\pi /3))]=\exp(-3\rho A/2).
\end{equation}

Since this gives the correct $N=2$ result if we replace 3 by 2 in the cosine in the first exponent, one might be tempted to generalize to all $N$ by using
$\cos(2\pi /N)$ in the exponent.  This would be correct in principle if there were only one kind of vortex, namely the one with $J=1$.  But this is likely to be wrong; for one thing, it gives no area law in the large-$N$ limit.  We have assumed for simplicity (see the Appendix for a more accurate discussion) that all vortices of whatever charge contribute equally.  If so, and if $N$ is odd, an elementary calculation yields:
\begin{equation}
\langle W \rangle = \exp-\{\frac{2\rho A}{N-1}[ 1-\cos (\frac{2\pi}{N}) + 
1-\cos (\frac{4\pi}{N}) + \dots + 1-\cos(\frac{2\pi}{N}(\frac{N-1}{2})\} .
\end{equation}
Each term $1-\cos(2\pi J/N)$ represents the contribution of a vortex and and antivortex.  The sum over these terms is elementary, and yields:
\begin{equation}
\langle W \rangle = \exp-\{ \frac{\rho AN}{N-1}\} ;\;K_F(N) = \frac{\rho N}
{N-1}.
\end{equation}
A similar calculation for $N$ even, left to the reader, gives the same result.
Of course, (22) agrees with the previous answer for $N=2,3$.

Now we go on to the more difficult case of the adjoint potential.

\bigskip

\begin{center}
{\bf Adjoint Wilson Loop}
\end{center}

We begin with a theorem which follows from generalizing the explicit $SU(2)$ calculation given below to $SU(N)$, using some simple properties of the adjoint representation of the vortex fluxes $Q_J$ and techniques similar to those used for the fundamental representation above.  These properties, given in Section
IIB, are that $Q_J$ has $(N-1)^2$ eigenvalues of 0, $N-1$ eigenvalues of +1, and $N-1$ eigenvalues of -1.  It then turns out (we leave details to the reader) that for any $N$ the leading term in the $\epsilon$ expansion of the adjoint potential $V_A(R;N)$, where $R$ is the separation between the two long sides of an adjoint Wilson loop, is a universal function
independent of $N$, when expressed in terms of the fundamental string tension and the mass:
\begin{equation}
V_A(R;N) = V_A(R;2)=\frac{K_F}{M}U(MR)(1+O(\epsilon )).
\end{equation}
So we need only calculate the universal function $U(R)$ for $SU(2)$.

Because all the eigenvalues of $Q_J$ are integral, and because we explicitly show periodicity in the vortex flux, there is no contribution to the adjoint potential from the long-range pure-gauge part of the vortex potential as given in equation (2).  It is useful, in fact, to define a short-ranged Abelian gauge potential, which is all that will appear in the adjoint potential:
\begin{equation}
A_i(x) = 2\pi \epsilon_{ij}\partial_j\Delta_M(x).
\end{equation}
Its flux tends to zero as the surface defining the flux tends to infinity.  

Write the adjoint Wilson-loop potential in $SU(2)$ for the dilute-vortex model as:
\begin{eqnarray}
\langle W_A(R)\rangle = \exp(-TV_A(R)) = \\ 
\frac{1}{Z}{\large \{}\sum \frac{1}{K!} \sum_1 \cdots \sum_K TrP\exp(i\int d\tau
J_a \dot{x}(\tau)\cdot A^a[x(\tau)]{\large \}} \nonumber
\end{eqnarray}
with the trace and group generators $J_a$ in the adjoint representation.
Here $T$ is the length of the long sides of a Wilson rectangle, and $R$ the length of the short sides.
The path-ordering prescription affects only the generators, as expressed in the formula
\begin{equation}
P(J^aJ^b\dots ) = \sum_{perm} J^aJ^b\dots \Theta(\tau_a \geq \tau_b\geq  \dots )
\end{equation} where $\Theta$ is one if the $\tau$-variables are ordered as shown, and zero otherwise; the sum is over all permutations of the indices.

We first show that in the leading cumulant term, found from the $K=0,1$ terms of (25), the path-ordering prescription can be ignored.  We then show that this leading-order term exponentiates when higher values of $K$ are considered, leaving a residual term of $O(\epsilon)$ and higher.  Path ordering is important in this residual term.  A proper cluster expansion emerges, in which all terms of $\log \langle W \rangle $ are linear in $T$ as $T$ approaches infinity.
If for the moment we accept this, then we can summarize our results for the leading, or one-vortex, term by saying that one can replace the actual group integral, in any representation $R$ of any $SU(N)$, by a discrete average.
In the $K=1$ sector there is only a single vortex, whose collective coordinates we indicate by the subscript 1.  Then we claim:
\begin{eqnarray}
\sum_1 TrP\exp (i\int d\tau
J_a \dot{x}(\tau )\cdot A^a[x(\tau );a_1,u_1]= \\
\rho \int d^2a_1 \frac{1}{D_R}\{  \sum \exp (i\hat{Q}_{J,R}\int d\tau \dot{x}(\tau )\cdot 
A[x(\tau )-a_1])\} .\nonumber
\end{eqnarray}
Here the $SU(2)$ gauge potential is related to the Abelian gauge potential by 
\begin{equation}
A_i^a(x-a;u)=\hat{e}^a(u)A_i(x-a)
\end{equation}
where on the right is the Abelian potential of (24), and $\hat{e}^a(u)$ is a unit vector.  Actually, this vector depends not on the full group variables, but only on the coset variables discussed in the Appendix.  For the group $SU(2)$ this coset is $SU(2)/U(1)$  and the unit vector just depends on the usual polar angles.  The $\hat{Q}_{J,R}$ are the eigenvalues of the vortex flux matrix $Q_J$ in the representation $R$, and $D_R$ is the dimension of this representation.  Note, by the way, that replacing the group integral by the above discrete average also yields the results of Section IVA for the fundamental Wilson loop.

The proof of this formula is simple.  One sees from the above that the group-generator term reduces to $Tr(\hat{e}\cdot J)^N)$ in the Nth order term of the expansion of the single-vortex path-ordered product.  Only even $N$ contributes, and for the adjoint of $SU(2)$ this reduces to  $Tr(\hat{e}\cdot J)^2)$, because   $ (\hat{e}\cdot J)^3=\hat{e}\cdot J$.  This trace is just 2, independent of the ordering of the generators in the original expression.

Using equation (27) in equation (25), one easily finds the $K=0,1$ contribution to the adjoint Wilson loop:
\begin{equation}
\langle W \rangle_{0,1}=3\exp \{ -\frac{2}{3} \rho \int d^2a[1-\cos\oint dx
\cdot A(x)] \} .
\end{equation}
As claimed, it is periodic in the (Abelian) flux.  It would be a serious mistake to expand the cosine, saving only quadratic terms, as one would do for a Gaussian distribution.  

It only remains to calculate the translational collective-coordinate integral.  
This has three terms: One from the vortices outside the Wilson loop to the left; an equal term for vortices outside to the right; and one for those inside.  By inside and outside we refer to the collective coordinates; since the vortices themselves are fat, they overlap the Wilson loop if they are within a distance $1/M$.  In fact, these are the only vortices which can affect the adjoint potential, which is really like a perimeter term in that only vortices near the perimeter can contribute.  This collective-coordinate integral was done some time ago \cite{c83}, and the result for the adjoint potential (using (22) to express
$\rho$ in terms of $K_F$):
\begin{eqnarray}
V_A(R)=\frac{K_F}{3M}\{ 2\int_0^{\infty} dy[1-\cos (\pi e^{-y}-\pi e^{-(y+MR)})]+ \nonumber \\
+\int_0^{MR}dy[1-\cos (\pi e^{-y}+\pi e^{y-MR})] \}
\equiv \frac{K_F}{M}U(MR) 
\end{eqnarray}
The calculation previously cited \cite{c83} of this integral was done for $SU(3)$, and it has exactly the same form, in accordance with our previously-stated theorem.

In Fig. 1 we show a plot of the potential $U(MR)$.  It has a more-or-less linearly-rising term for a distance of order $1/M$, and then it settles down to a constant.  The asymptotic value $V_A(\infty )$is about 2.2$K_F/M$, which should be comparable to $2M$, yielding $M\simeq 1.1K_F^{1/2}$, or about 460 MeV using the usual value for the $SU(3)$ string tension.  The slope of the linearly-rising term is about 1.5 $K_F$, not as big as Casimir scaling would suggest, but we see that there is nothing in the underlying physics to suggest that Casimir scaling should hold anyway.  These numbers are in any case not very accurate, first because the asymptotic value is only roughly 2$M$, and second because there are other contributions from, {\it e.g.}, instantons.  

\bigskip

\begin{center}
{\bf C.  Finite-density corrections to the adjoint potential}
\end{center}

One must first show that in the two-vortex sector, which has terms of $O(T^2)$ as well as $O(T)$, these quadratic terms cancel when $\log \langle W\rangle$ is formed.  We form in the usual way the cumulant through two-vortex terms, arriving at:
\begin{eqnarray}
-\log \langle W \rangle = \frac{2}{3}\sum_1 (1-\cos \psi_1) -\\ \nonumber
-\frac{1}{3\cdot 2!}\sum_1\sum_2\{ TrP\exp i\oint dx_iJ^aA^a(x;c_1,c_2)- \\ \nonumber
-\frac{1}{3}[1+2\cos \psi_1 +2\cos \psi_2 + 4\cos \psi_1 \cos \psi_2]\}.
\end{eqnarray}
Here the subscripts 1,2 refer to the collective coordinates (see (15)), and
\begin{equation} 
\psi_1 = \oint dx_i A_i(x-a_1;u_1);\;A^a_i(x;c_1,c_2) = \hat{e}^a(u_1)A_i(x-a_1) + (1 \leftrightarrow 2).
\end{equation}
There are now two group integrations and two unit vectors, so the trace of a product of generators times these unit vectors is not so simple, and path-ordering is important, at least at higher than quadratic order (in $A^a$ of equation (32)).  

Construct the usual series expansion of the path-ordered product, in which only even-order terms need be kept.  At 2Nth order one encounters terms like
\begin{equation}
TrP(\hat{e}^a(u_1)J^a\hat{e}^b(u_2)J^b\hat{e}^c(u_1)J^c\dots )
\end{equation}
in which there are $R$ terms in $\hat{e}^a(u_1)J^a$ and $N-R$ terms in
$\hat{e}^b(u_2)J^b$.  These come in all permutations, so it is not possible to gather terms in vortex 1 separately from those referring to vortex 2 without paying attention to the fact that the generators do not commute.  However, let us proceed by replacing (33) by
\begin{equation}
Tr[(\hat{e}^a(u_1)J^a)^R(\hat{e}^b(u_2)J^b)^{2N-R})
\end{equation}
plus a remainder which reinstates the correct expression.  The trace in (34) is now elementary; if $R\neq 0,2N$ one reduces it, as before, to
\begin{equation}
Tr[(\hat{e}^a(u_1)J^a)^2(\hat{e}^b(u_2)J^b)^2]. 
\end{equation}
One can now integrate over the group coset, replacing $\hat{e}_i(u_1)\hat{e}_j(u_1)$ by $(1/3)\delta{ij}$.  Then (34) reduces to 
$(1/9)Tr(J^2)^2=4/3$.  But if $R=0,2N$ the trace becomes $2/3$.  It is clear that the terms in which (33) has been used reduces to the sum
\begin{eqnarray}
3+\sum_{N=1} \frac{(-)^N}{(2N)!}\sum_{R=0,even}\psi_1^R\psi_2^{2N-R}\frac{(2N)!}
{R!(2N-R)!}\times \\ 
\times [\frac{4}{3}(1-\delta_{R,0}-\delta_{R,2N})+\frac{2}{3}(\delta_{R,0}+
\delta_{R,2N})]. \nonumber
\end{eqnarray}
This sum is easily done, and it completely cancels the third term in brackets on the right-hand side of (31).  This cancellation, of course, is necessary for proper clustering, in which $\log \langle W \rangle$ must be linear in $T$.

This leaves only the remainder term, which we will discuss explicitly only in
the lowest order, namely, $O(A_1^2A_2^2)$.  At this order one encounters only two separate values for the trace, which is a trace of four group generators.  In sixteen of the twenty-four terms in the path-ordered product we find the value given in (35), while the remaining eight traces are only half as big.
The result is that the true two-vortex term in the cumulant expansion is
\begin{equation}
\frac{1}{3\cdot 4! \cdot 2!}(-18)(\frac{2}{9})\int d\tau_1 \dots d\tau_4 F(\tau_1-\tau_2)F(\tau_3-\tau_4)[\Theta(\tau_1 \geq \tau_3 \geq \tau_2 \geq \tau_4) + \dots ] . 
\end{equation}
Here
\begin{equation}
F(\tau_1-\tau_2)=\rho \int d^2a \dot{x}(\tau_1)\cdot A(x(\tau_1)-a)
\dot{x}(\tau_2)\cdot A(x(\tau_2)-a)
\end{equation}
is constructed from the lowest-order semi-classical propagator of the vortices, and in (37) the ellipses in the square brackets indicate seven other permutations.  We need not write these explicitly, since they all give the same result; these other permutations are those generated by all exchanges of the $\tau$s under which $F(\tau_1-\tau_2)F(\tau_3-\tau_4)$ is invariant. One can readily verify that because of the $\Theta$-function in (37) this term is $O(T)$, and not $O(T^2)$ as it would be without the $\tau$-ordering.

Because this term is only the first term of an infinite expansion, it makes no particular sense to evaluate it any further.  It is, as advertised, of $O(\rho/M^2)$ compared to the leading term as given in (29).

\bigskip

\begin{center}
{\bf V.  THE CENTER-VORTEX MODEL, d=2 QCD, AND LARGE-N FACTORIZATION}
\end{center}

In two dimensions, QCD without fermions\footnote{With fermions, $d=2$ QCD can show screening in any representation if the fermions are massless; see Ref. \cite{avs}.)} is an exactly-soluble theory (see, {\it e. g.}, \cite{kk}).  All representations are confined, and there is Casimir scaling.  The theory is simply one of free massless propagators coupled in the usual way to Wilson loops.  Here we discuss how the dilute center-vortex picture in $d=2$ resembles, and differs from, QCD in this dimension.

Let us construct the gauge propagator from the collective fields of the vortex condensate in the usual way.  This propagator is:
\begin{equation}
\langle A^a_i(x)A^b_j(y) \rangle = \frac{\delta_{ab}}{N^2-1}\sum_{a,u}
\sum_{J=1}^{[N/2]}(-2)Tr(Q_J^2)A_i(x;a,u;J)A_j(y;a,u;J)
\end{equation}
Here the gauge potentials are the vortex potentials, as functions of their collective coordinates, as given in equations (2,3), and $\sum_{a,u}$ is the integral over these collective coordinates.  To mimic $d=2$ QCD in the 
{\em fundamental} representation it is enough, as we have already done above, to save only the pure-gauge long-range part of the vortex, that is, the $\Delta_0$ piece in (2).  (Saving the massive part would give rise to perimeter-law corrections not found in $d=2$ QCD.)  The sum over collective coordinates has the form already used:
\begin{equation}
\sum_{a,u}=\frac{\rho}{N-1}\int d^2a \int d(u)
\end{equation}
(recall that the group integration is normalized to unity).  A quick calculation shows that the collective propagator coming from the long-range pure-gauge part is 
\begin{equation}
\langle A^a_i(x)A^b_j(y) \rangle = \frac{\delta_{ab}(2\pi )^2\rho}{3(N-1)g^2}\Delta_{ij}(x-y)
\end{equation}
where
\begin{equation}
\Delta_{ij}(x-y)=\frac{1}{(2\pi )^2}\int d^2k(\delta_{ij}-k_ik_j/k^2) \frac{e^{ik\cdot x}}{k^2}
\end{equation}
is the gauge propagator of $d=2$ QCD.  We then need only require that
\begin{equation}
\rho = \frac{3(N-1)g^2}{4\pi^2}
\end{equation}
to recover $d=2$ QCD exactly.  (Note that this requirement survives the large-$N$ limit).

However, we {\em cannot} do the same for the adjoint representation which, as we have shown, is not sensitive to the long-range pure-gauge part of the vortices.
The adjoint potential in the center-vortex model is very different from its $d=2$ counterpart.

It is difficult to reconcile this view of the center-vortex picture with the large-$N$ factorization property (see, {\it e.g.}, ref. \cite{gh}) which leads to Casimir scaling and an adjoint area law.  Factorization begins with the identity
\begin{equation}
Tr_A U = Tr_FU Tr_FU^{\dag} -1;\;U=\exp g\oint dx_iA_i(x),
\end{equation}
(where the superscript $A$ refers to the adjoint representation, and $F$ refers to the fundamental)
followed by the large-$N$ prescription $\langle TrUTrU^{\dag} \rangle \linebreak  \rightarrow \langle TrU \rangle \langle TrU^{\dag} \rangle + O(1/N^2)$.
This second step is certainly true in both true large-$N$ QCD and in the truncated version of the center-vortex model discussed immediately above. 
However, when the full center-vortex model is used one {\em cannot} simply apply the perturbative rules of factorization at large $N$.  Instead, one sees that, at any finite $N$, a vortex linked (unlinked) to the (large) Wilson loop $U$ is also linked (unlinked) to the loop $U^{\dag}$.  The long-range pure-gauge part of the vortex supplies equal and opposite phase factors from the center of the group to $U$ and $U^{\dag}$.  These phase factors are multiples of the identity, and can be pulled outside the traces; they cancel in the product in (44), and cannot lead to an area law after averaging over the vortices.  However, the short-range vortex contributions contribute factors which are not multiples of the identity.  After averaging, these give the perimeter law we have calculated in earlier sections.  In the full center-vortex model, there is no justification for using large-$N$ factorization, which asserts independence of the phase factors in $U$ and in $U^{\dag}$, in the specific context of equation (44) relating adjoint and fundamental representations.  There is, however, no problem in using large-$N$ factorization in other circumstances..  For example, the expectation value of a product of {\em distinct} Wilson loops is a product of expectation values of the individual loops, in leading order in $N$.

The above view, presenting a conflict between factorization and the behavior of the adjoint Wilson loop at large $N$, is by no means the only one possible.  For example, the authors of Ref. \cite{fgo2} argue that the vortex thickness grows at large $N$, perhaps like ln $N$, so that a hypothesized Casimir-scaling regime in the center of the vortex grows to fill any Wilson loop of fixed size, however large.  In this case confinement at large $N$ would be completely different from the averaging over group-center phases which the vortex model shows at finite
$N$.      
\bigskip

\begin{center}
{\bf VI.  SUMMARY AND CONCLUSIONS}
\end{center}

We have shown how to calculate the adjoint potential in the center-vortex picture; all the explicit work was done in $d=2$.  The result was that the adjoint potential is a universal (for all $N$) function of the form $(K_F/M)U(MR)$, where $U$ shows a roughly-linear regime but then asymptotes to a constant value, representing string breaking when about $2M$ of energy is stored in the adjoint string.  There is no particular relation between the slope of the linear adjoint potential and the
fundamental string tension.  To the extent that our calculations apply at least qualitatively in $d=3,4$ there are other contributions showing the same general structure which should be evaluated, {\it e.g.}, the instanton contribution (instantons are short-ranged \cite{c79} like the adjoint vortices), which further obscure any relation like Casimir scaling between fundamental and adjoint Wilson loops.

Even though we worked mostly in $d=2$ we emphasized that the center-vortex picture in this dimension is not the same as $d=2$ QCD, although it can be made to look the same for the fundamental representation.  In particular, we were interested in the large-$N$ limit, and had to show that vortices of large ($J\sim N$) flux, corresponding to elements of the center group far from $J=1$, could have free energies which scaled appropriately so as to contribute to the fundamental string tension.  Correct large-$N$ scaling occurs also if all vortices have the same free energy, which we have to hope is a dynamical requirement of QCD in $d=3,4$.  In any event, we assumed this equality of free energies when discussing general properties of the center-vortex picture.

Another distinction between the large-$N$ center-vortex picture and true $d=2$   QCD is that conventional factorization of matrix elements does not occur in the center-vortex picture.  This is because overall phase factors associated with the center of the group, and which give fundamental-representation area laws, cancel in the formula (44) to which factorization is applied.    

What of this survives in the physically-interesting dimensions for the center-vortex model, $d=3,4$?  We believe that the qualitative features survive:  There is a smooth large-$N$ limit for the string tension in the fundamental representation, and a universal $N$-independent form for the adjoint potential at low densities, when expressed in terms of the string tension and the vortex size (or gluon mass).  This potential rises more or less linearly, but its slope is not necessarily related by Casimir scaling to the fundamental string tension.  Of course, one should expect that the {\em total} adjoint potential will depend on the dimensionality, because the various other contributions to this potential certainly depend on it, and the center-vortex contribution by itself should depend on dimension.  It would be interesting to make lattice simulations with a lattice action that surpressed all but the center vortices, just to see how close these come to yielding the adjoint potential (apart from perturbative one-gluon exchanges, etc.).  

There should be numerous other tests of the center-vortex picture, going well beyond the present test via the fundamental string tension.  For example, it has been shown \cite{c96} that the center-vortex picture prescribes a triangle law for the forces between quarks in an $SU(3)$ baryon, rather than the so-called Y-law.  One should also attempt to calculate effects coming from the merging and splitting of vortices of different charges, as described in Sec. IIB \cite{c79,c94}, with fractional Chern-Simons numbers associated with these vortex vertices; an example would be the estimation of the topological susceptibility, or the response to a $\theta$-term in the Lagrangian.  Of considerable interest for future work is understanding the effects of vortices'
merging at a point ($d=3$) or line ($d=4$), which in three dimensions is associated with generation of fractional Chern-Simons number and in four dimensions with the response to a $\theta$-term in the Lagrangian.

\bigskip

\begin{center}
{\bf ACKNOWLEDGMENTS}
\end{center}
I thank J. Greensite for valuable correspondence concerning the large-$N$ properties of center-vortex models.
This work was supported in part by the National Science Foundation under 
Grant PHY-9531023.

\newpage

\begin{center}
{\bf APPENDIX}
\end{center}

As discussed in the text, vortices of charge $J\;(J=1,\dots [N/2])$ have different actions $I(N,J)$, so that it is not clear that all elements of the center group are treated equally, especially at large $N$.  (There is no problem at $N=$2, 3, where there is only one value of the vortex action.)
Considered naively, this poses severe problems for the existence of a large-$N$ limit; for example, we have already pointed out in connection with equation (21) that if only the vortex with lowest action---the $J=1$ vortex---is saved, the fundamental string tension vanishes at large $N$.  The action of the vortex of flux $J$ is proportional to $J(N-J)/Ng^2$ which behaves like $N^2$ when $J\sim N$, so that the exponential of the action would seem to vanish very rapidly, and in general the string tension would indeed vanish in large $N$.  Here we show that in $d=2$ one can produce a viable large-$N$ limit by (1) imposing a single condition on some parameters of the $d=2$ center-vortex model; (2) choosing correctly some non-leading terms in the dependence of the coupling constant $g^2$ on $N$ (these are not the same non-leading terms that are found in any particular theory, such as $d=2$ QCD).  In the dimensions where there is supposed to be a center-vortex dynamics produced by the underlying QCD theory, that is, $d=3,4$, such conditions cannot be imposed by hand, as we do in $d=2$, but must follow from the underlying theory.  This is a very difficult problem, and we do not address it here.  Our only concern is whether a $d=2$ center-vortex model can be tuned (not fine-tuned; no large or small numbers appear) to have a sensible large-$N$ limit.    

Consider first the partition function, which can be written (expanding somewhat the condensed notation of equation (14))
\begin{equation}                                        
Z^{1/2}=\sum_1 \frac{1}{K_1!}\cdots \sum_{[N/2]}\frac{1}{K_{[N/2]}!}= \exp
\{ \sum_1 + \cdots \sum_{[N/2]} \} .
\end{equation}
Here the sum labeled 1 goes over the collective coordinates of vortices with charge $J=1$, etc.  By terminating the sum at $J=N/2$ (at large $N$ we need not distinguish even and odd $N$, so the brackets indicating integer part can be dropped), we include only the vortices with positive charge; squaring this expression takes into account the equal contribution of the anti-vortices.

The specific meaning of the collective-coordinate sum for vortex $J$ is:
\begin{equation}                                      
\sum = \frac{const.J(N-J)}{Ng^2}\int d^2a \int_{C(J,N)} d(u) (\beta/g)^{\nu(N,J)}e^{-I(N,J)}
\end{equation}
where the translation coordinates are $a$, the group coordinates are $u$ (the group integration is over a coset $C(J,N)$ defined below, and is not normalized to unity, as in the main text), $\nu(N,J)$ is the number of group zero modes, and $I(N,J)$ is the action of the vortex of charge $J$.  The constant $\beta$, coming from the group zero-mode normalization, is not a function of $N,J$, as one easily checks.  We have explicitly displayed the factors associated with the translational zero modes; presumably there is no conformal mode, because of the presence of a mass in the vortex solution.
Note that in $d=3,4$ there would also be integrals over configurational degrees of freedom of the string or closed surface.
We choose the scale of mass so that the gluon mass, or vortex inverse size, is unity.

There will be a smooth large-$N$ limit if, with
\begin{equation}                                  
Ng^2 = c^2 + O(1/N)
\end{equation}
where $c$ is a constant independent of $N,J$, the partition function and various expectation values exist at $N= \infty$.  In particular, the sum over $J$ in the second equation of (45) must have a smooth limit.

The group integration runs over the  parameters of a coset which is $SU(N)$ divided by the invariance subgroup of the vortex.  From the explicit representation of the vortex flux matrix $Q_J$ of Section II, we know that this coset is 
\begin{equation}                                   
\frac{SU(N)}{SU(N-J) \otimes SU(J) \otimes U(1)}.
\end{equation}
The $U(1)$ here is essentially generated by $Q_J$ itself, except that the range of the angular parameter multiplying the generator is not $2\pi$, but
$2\pi (2N(N-J)/J)^{1/2}$ (see Bernard \cite{cb79}).  This number is the volume $V_{J,N}(1)$ of the $U(1)$ subgroup.  Note that the number of group zero modes is just:
\begin{equation}                                      
\nu (N,J) = N^2-1 - [(N-J)^2-1] -[J^2-1] -1 = 2J(N-J).
\end{equation}
  With the usual normalization of group generators ($Tr[(\lambda_a/2)(\lambda_b/2)]=(1/2)\delta_{ab}$) the coset volume can be calculated from the well-known (see, {\it e.g.}, \cite{cb79}) volume $V(N)$ of the group $SU(N)$:
\begin{equation}                                                 
V(N) = N^{1/2}2^{-(N-1)/2}(4\pi )^{(N-1)(N+2)/2}\prod_{r=1}^{N-1}
\frac{1}{r!}.
\end{equation}
The needed coset volume is
\begin{equation}                                                  
V(J,N)= \frac{V(N)}{V(N-J)V(J)V_{N,J}(1)}.
\end{equation}

Next, turn to the action factor $\exp(-I(N,J))$, which we write in terms of a positive constant $\alpha$, independent of $J,N$:
\begin{equation} \exp (-I(N,J))=\exp (\frac{-\alpha J(N-J)}{Ng^2})     
=\exp (\frac{-\alpha J(N-J)}{c^2}).
\end{equation}
The last factor we need comes from the zero-mode normalizations:
\begin{equation}                                                   
(\frac{\beta}{g})^{2J(N-J)}= (\frac{\beta^2}{c^2} N)^{J(N-J)}.
\end{equation}
We have explicitly written only the leading-order dependence of $g^2$ on $N$.

Write the collective-coordinate integral (46) as:
\begin{equation} \sum = \int d^2a R(J,N);\;R(J,N)=\frac{J(N-J)}{Ng^2}V(J,N)[\frac{\beta^2N}{c^2}e^{-\alpha / c^2}]^{J(N-J)}.
\end{equation}
The partition function (or expectation values) depends on the (weighted, if an expectation value) sum over $J$ of $R(N,J)$, as in the second equation in (45) expressing the partition function.  Because $N$ is large, we can write such sums as integrals over a variable $x\equiv J/N$.  For example, the fundamental Wilson loop expectation value can be written ({\it cf.} equation (21)):
\begin{equation}
-\log \langle W \rangle = \frac{2N^2A}{g^2}\int_0^{1/2} dxx(1-x)[1-\cos (2\pi x)]
R(Nx,N).
\end{equation}
Here $A=\int d^2a\Theta_W(a)$ is the area of the loop ({\it cf.} equation (4)).

By examining various terms in $R$, one finds the generic behavior:
\begin{equation} R=\exp [h(x)N^2\log N+ i(x)N^2 + j(x)N\log N + k(x)N +l(x)\log N \cdots ]
\end{equation}
The functions $h,i,j,k\dots$ can be found with the aid of Stirling's formula and the Euler-Maclaurin sum formula, and one discovers that $h$ vanishes identically.  This is essential; if it did not vanish identically, it could not be tuned away, because none of the terms in $R(J,N)$ that depend on the various parameters we have introduced appear in the function $h$. They only appear in less-singular terms. 

The next-leading term is $i(x)$:
\begin{equation}
i(x)=x(1-x)[\log (\frac{4\pi \beta^2}{c^2})+\frac{3}{2} -\frac{\alpha}{c^2}]
+\frac{1}{2}[x^2\log x+ (1-x)^2\log (1-x)].
\end{equation}

We now find that we can make $i(x)$ vanish at its upper limit of 1/2 provided that we choose
\begin{equation}
\frac{2\pi \beta^2}{c^2}\exp[\frac{3}{2}-\frac{\alpha}{c^2}]= 1.
\end{equation}
The significance of this is that if we ignore for the moment all the terms sub-leading to $i(x)$ the integral in the Wilson-loop formula (55) is $O(1/N)$, and not $O(\exp (-N^2))$.  

It remains to deal with the next-leading terms.  It turns out that $j(x)$ in (56) also vanishes identically.  The next-leading, or $O(N)$ term, can be rendered harmless by choosing the correct coefficient for $1/N$ corrections to the large-$N$ scaling of the coupling constant, as in (47). Ultimately further corrections to the coupling constant can be tuned to give a non-vanishing fundamental string tension.  But for the adjoint representation one may not approximate the sum over vortices by an integral, since the factor $1-\cos (2\pi x)$ in (55) is replaced by $1-\cos (2\pi J )\equiv 0$, plus, of course, the perimeter terms we dealt with in the main text.    

This is, of course, all done by hand in the $d=2$ center-vortex model, and it is a hope that gluon dynamics in higher dimensions achieves the same result.  It is worth emphasizing that achieving correct large-$N$ behavior of the fundamental string tension requires going beyond leading order in $N$.

\newpage
\begin{center}
{\Large \bf Figure Captions}
\end{center}

\bigskip

\noindent 1.  Plot of the adjoint potential $U$ vs. MR.
\newpage
\epsfig{file=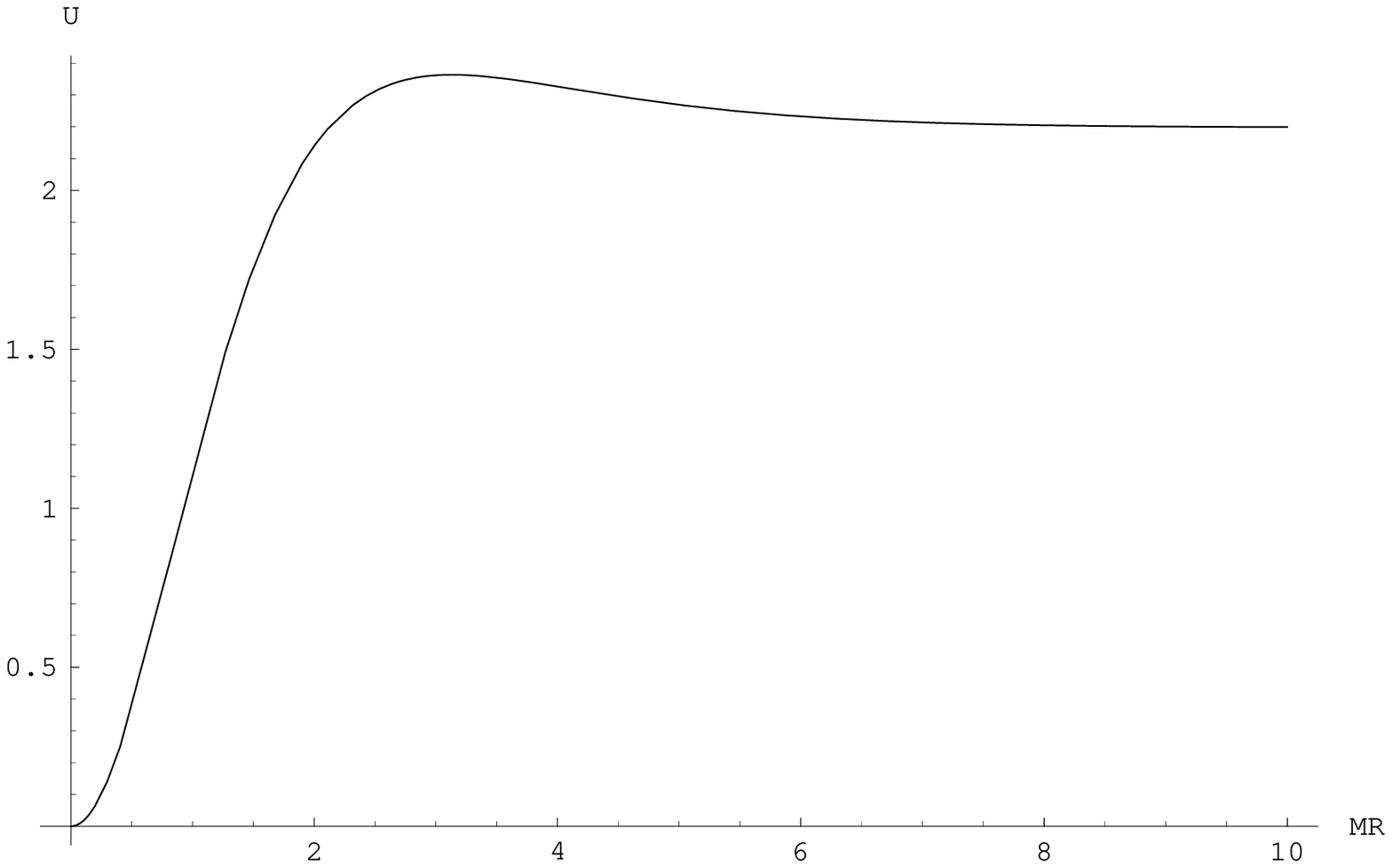,clip=} 

\end{document}